# Interactive effects of multiple stressors in coastal ecosystems

**Shubham Krishna[1*]** | **Carsten Lemmen[1]** | **Serra Örey[2,3]** | **Jennifer Rehren[3]** | **Julien Di Pane[4]** | **Moritz Mathis[1]** | **Miriam Püts[3]** | **Sascha Hokamp[5]** | **Himansu Kesari Pradhan[1]** | **Matthias Hasenbein[6]** | **Jürgen Scheffran[5]** | **Kai W. Wirtz[1]**

[1]Institute of Coastal Systems - Analysis and Modeling, Helmholtz-Zentrum Hereon, Geesthacht, 21502, Germany

[2]Bremerhaven University of Applied Sciences, Bremerhaven, 27568, Germany

[3]Johann Heinrich von Thünen-Institut, Bremerhaven, 27572, Germany

[4]Electricité de France EDF, Paris, 75008, France

[5]Universität Hamburg, Hamburg, 20144, Germany

[6]Bundesamt Für Seeschifffahrt Und Hydrographie BSH, Hamburg, 20359, Germany

**Correspondence**
Dr. Shubham Krishna, Institute of Coastal Systems - Analysis and Modeling, Helmholtz-Zentrum Hereon, Geesthacht, 21502, Germany
Email: shubham.krishna@hereon.de

**Funding information**
Multiple Stressors on North Sea Life (MuSSeL, grant number: 03F0862A)

Coastal ecosystems are increasingly experiencing anthropogenic pressures such as climate heating, $CO_2$ increase, metal and organic pollution, overfishing and resource extraction. Some resulting stressors are more direct like fisheries, others more indirect like ocean acidification, yet they jointly affect marine biota, communities and entire ecosystems. While single-stressor effects have been widely investigated, the interactive effects of multiple stressors on ecosystems are less researched. In this study, we review the literature on multiple stressors and their interactive effects in coastal environments across organisms. We classify the interactions into three categories: synergistic, additive, and antagonistic. We found phytoplankton and mollusks to be the most studied taxonomic groups. The stressor combinations of climate warming, ocean acidification, eutrophication, and metal pollution are the most critical for coastal ecosystems as they exacerbate adverse effects on physiological traits such as growth rate, basal respiration, and size. Phytoplankton appears to be most sensitive to interactions between metal and nutrient pollution. In nutrient-enriched environments, the presence of metals considerably affects the uptake of nutrients, and increases respiration costs and toxin production in phytoplankton. For mollusks, warming and low pH are the most lethal stressors. The combined effect of heat stress and ocean acidification leads to decreased growth rate, shell size, and acid-base regulation capacity in mollusks. However, for a holistic understanding of how coastal food webs will evolve with ongoing changes, we suggest more research on ecosystem-level responses. This can be achieved by combining in-situ observations from controlled environments (e.g. mesocosm experiments) with modelling approaches.

## 1 | INTRODUCTION

Coastal ecosystems are exposed to a plethora of direct and indirect anthropogenic stressors such as climate heating, eutrophication, metal pollution, hypoxia, pH and salinity changes, and overfishing [1, 2, 3, 4]. These stressors are inducing serious and irreversible changes in marine and coastal food webs [5], and they do not act in isolation but instead simultaneously [3]. It has been suggested that by 2050 about 90% of the global ocean will be impacted by exposure to multiple stressors [6]. Interactions between stressors trigger responses in species and communities that are often different from the effects of the individual stressors [2, 7].

To understand the changes in coastal ecosystems and to improve marine and coastal management, a better knowledge of stressor interactions is required [4]. The net response to multiple stressors could be the sum of the individual responses, more than that, or less. When the combined effect to multiple stressors is equal to their expected cumulative sum, such a effect is called additive. When this is not the case, it is termed multiplicative or non-additive [8]. A multiplicative effect could, in turn, be synergistic or antagonistic. When the net effect of interactions is greater than the sum of individual stress responses, the response is called synergistic [2, 7]. On the contrary, in an antagonistic interaction, the effect intensity of the combined response is less than that of the cumulative single-stressor effects.

Occurrences of additive and multiplicative interactions have been reported in coastal ecosystems [2, 9]. For example, synergistic responses to elevated nutrient concentrations and metal loadings have been found in diverse plankton groups [10, 11, 12, 13]. Negative synergistic effects have been reported in benthic organisms (such as crabs and mollusks) on exposure to ocean acidification (OA), salinity, temperature, and metal pollution[14, 15, 16]. The combined effects of heat stress and ocean acidification have been far more adverse than their individual effects in a variety of mussel species [17, 18]. Likewise, the adverse effects of overfishing can be intensified by climate warming in top predatory fish species [19]. Stressor interactions also lead to antagonistic responses in coastal ecosystems. For example, it has been shown that elevated $CO_2$ resulted in higher chlorophyll biomass under low irradiance, thereby compensating for light stress [20, 21]. Similarly, low pH has been shown to have mitigated the negative effect of high temperature on egg volume in an estuarine fiddler crab [22].

In spite of the significant role of stressor interactions in driving ecosystem and community-level responses, investigations of these effects are largely disconnected from the implementation of conservation and management policies for coastal systems [5]. Over the last decades, research and management focussed on eutrophication problems in coastal environments [23]. The non-additive effects of eutrophication with other stressors such as climate warming, pollutants, hypoxia and changes in salinity have posed a multidimensional problem [24, 6, 25]. First, the intricacy of interactive effects makes it very difficult to elucidate the net response of an ecosystem in ever-changing environments [5]. Secondly, the occurrence of several stressors triggers differential responses across organisms. Thirdly, critical stressor combinations vary between taxonomic groups.

In this study, we address the aforementioned research gaps. With a focus on high-latitude coastal systems, we systematically review the existing studies on the interactive effects of multiple stressors in coastal habitats to answer the following important questions: 1) What are the critical stressors for triggering non-additive responses in coastal ecosystems? 2) What are the main stressor combinations for different taxonomic groups with respect to inducing negative synergistic effects? 3) What are the research gaps in multi-stressor studies? From our findings, we provide valuable insights to scientists and stakeholders with respect to the advancement of multi-stressor research and for management of coastal ecosystems.

## 2 | MATERIALS AND METHODS

A systematic literature search for scholarly articles relating to the topic was performed using Clarivate's Web of Science (WoS) Advanced search. The search included a topical search for multiple stressors, compound, cumulative or interactive effects at the coast or on the shelf, classified as an article or review. We implemented the following search query:
`"TS=((("multipl* stress*") OR ("compound* effect") OR ("cumulat* effect") OR ("interact* effect")) AND (coast* or shelves or shelf)) AND DT=("ARTICLE" OR "REVIEW")"`
Abstracts of all obtained records were prescreened independently by two researchers for false positives that should be discarded as not relating to the topic, e.g. articles relating to management and policy, society and socio-ecological systems, single stressors, freshwater systems, marshes or seismology. The review papers contained in the remaining records were screened for references to primary relevant literature that was not captured by the automated search. Citations within review papers that were identified as relevant after screening their abstract were included in the subsequent full-text analysis. All remaining and manually added full texts were distributed for detailed evaluation among the author team. We employed the Preferred Reporting Items for Systematic Reviews and Meta-Analyses (PRISMA, [26]) approach for our analysis. The systematic search, based on queries and keywords, on WoS yielded 814 papers that investigated interactive effects of multiple stressors in marine ecosystems. All of these papers were screened, first by an automated script and then manually, to select the studies that focused only on high-latitude coastal ecosystems, which reduced the number of papers from 814 to 400. Out of these, a few were review or synthesis papers which were discarded from the analysis, and the remaining were distributed amongst the co-authors to review and fill up the "Summary Table" (Table 1). In total, we could identify 198 studies in which non-additive or additive effects were reported. All further analyses were performed based on the information provided in this table.

To quantify the magnitude of interactive effects (synergistic = SYN, additive = ADD, and antagonistic = ANT) for stressor combinations, we implemented a fuzzy coding/scoring method where scores between 1 to 3 were assigned for the reported responses. For example, if a study reports only a synergistic effect for a given stressor combination, a score of 3 is assigned for SYN and a score of 0 for ANT and ADD. If two types of effects (e.g. SYN and ANT or ADD and SYN) are reported then a score of 1.5 is given to each and 0 to the third effect which is missing. And, if all three effects (SYN, ANT, and ADD) could be identified then a score of 1 is assigned to all. For a given stressor pair, the sum of all three effects is always 3 (SYN + ANT + ADD = 3). If there are more than two stressor combinations, then the same procedure is followed for the respective stressor pairs resulting from that particular combination (e.g. 4 stressor combination yields 6 stressor pairs). The final score for a given effect (SYN or ADD or ANT) is then calculated by adding up their individual scores identified from different stressor combinations. The detailed schematic of this scheme is illustrated in Figure 8.

## 3 | RESULTS

### 3.1 | Interactive effects in coastal oceans

Interactive effects are reported at different levels of the coastal ecosystem; the majority (n=109, 55%) of studies focussed on the species or organism level, 36% (n=71) on the community level, and 9% (n=20) on the entire ecosystem. The types of interaction effects reported can be classified as synergistic, antagonistic, or additive: Most of the studies (130 out of 200) reported synergistic or additive effects, indicative of compounding of responses to multiple stressors and that research focuses on such effects (Fig. 1). Out of those 132, 68 studies reported only SYN effects and 62 only ADD effects, and 15 reported both. Where as ANT effects are identified in only 20 studies (10%). A few reported both the ANT- and SYN-type responses and only 16 studies identified all three effects. Most frequent individual stressors in a multi-stressor constellation are temperature (Temp, 25%, n=50), followed by nutrient loading (Nut, 17%, n=33) and toxic metals (Metal) and ocean acidification (OA, together 28%, n=56) (Fig. 2). Sedimentation loads (Sed), hypoxia (DO) and turbulence (Turb) are least studied. Quantification of the interactive effects (SYN, ANT, and ADD) for the stressor combinations of the most reported stressors is shown in Figure 3. The combinations of acidification–eutrophication (OA and Nut) and metal pollution–eutrophication (Metal and Nut) have mostly been reported as synergistic (with >50% score), whereas acidification–metal pollution (OA and Metal) has been reported mostly as additive ($\sim$ 70% score). The Heating–acidification (Temp and OA) combination is expressed equally as additive, synergistic and antagonistic.

### 3.2 | Break-down of stressor interactions with respect to taxonomic groups

The most studied taxonomic groups, with respect to stressor interactions, are mollusks, phytoplankton, and seagrass. They contribute to 66% (24% for mollusks, 21% for phytoplankton, and 21% for seagrass) of all studied groups (Fig.4). The remaining third includes fish (10%), zooplankton (9%), corals (2%), and polychaetes (3%).

Next, we investigated what kind of interactive effects are reported for individual groups. Exposure to multiple stressors triggered synergistic responses mostly in phytoplankton and mollusks (Fig. 5). Most of these responses were identified in physiological traits such as growth rate, carbon fixation and respiration rates (Fig. 9). By contrast, all three effects (SYN, ADD, ANT) have been reported for fish and decapods, which prevents the identification of a clear prevalence of one effect type (Fig. 5). This difference could be partially attributed to research constraints arising from the complexity of studied organisms or systems. At higher body size the fraction of manipulative experiments decreases. For smaller organisms, such as phytoplankton, zooplankton, and mollusks, more than 50% of studies are performed in controlled laboratory setups, whereas this fraction falls below 50% for fish and decapods and below 20% for seagrass (Fig. 6). Larger organisms are preferably studied in their natural and quasi-natural habitats. Likewise, in-situ experiments are preferred for studying ecosystem-level effects of multiple stressors.

So far, we found that phytoplankton and mollusks are the most studied organisms and exposure to simultaneous stressors triggers synergistic responses in their physiological traits. The most critical stressor combinations for phytoplankton and mollusk species in coastal oceans were identified by a fuzzy coding approach, where stressor combinations were given scores for synergism. For phytoplankton, the metal-nutrient (Metal and Nut) combination got the highest score (Fig. 7), as the exposure to nutrient and metal pollution usually instigates synergistic responses at physiological level in autotrophs. Likewise, the temperature-acidification (Temp and OA) stressor pair got the highest score for mollusks (Fig. 7), indicating that the combination of warming and ocean acidification is likely to synergistically affect their physiological rates.

# 4 | DISCUSSION

## 4.1 | Critical stressor combinations for coastal food webs

We identified OA, eutrophication, and metal pollution as the most critical stressor combinations for coastal ecosystems. All of these stressors directly follow from human activities on land that strongly influence coastal environments [27, 28]. Eutrophication and metal pollution negatively affect the ecological health of coastal systems in isolation and their interactive effects are often even more adverse [29, 30, 31, 32, 33]. For example, several sites in the Bohai Sea (China) have been identified as high-risk areas for autotrophs, crustacean, fish and mollusks due to the compounding effects of the elevated nutrient and metal concentrations [33]. For phytoplankton, the metal/nutrient stressor combination appears to be most critical and is discussed below.

## 4.2 | Phytoplankton

A full range of physiological effects can result from the interaction of eutrophication and metal pollution in phytoplankton, ranging from acute toxicity to sub-lethal or positive effects. In addition, shifts in community structure have been observed [34]. Toxic metals have inhibitory effects on germination of phytoplankton in eutrophic coastal waters [35]. It has been reported that the combined effect of metal and nutrient stressors results in significant changes in phytoplankton community structure leading to a shift from bigger diatoms to smaller and harmful cyanobacteria and dinoflagellates [36, 37]. Metals, such as copper and cadmium, inhibit the growth of large algal species (e.g. diatom), even in nutrient-replete conditions, by suppressing the uptake of enzymes and causing cell leakage. The shift to harmful algal species results in a reduction of food quality for secondary producers and thereby affects the trophodynamics in the coastal food web. Nutrient enrichment in many coastal waters can considerably enhance trace metal uptake in phytoplankton, which leads to bioaccumulation of metals at the higher trophic levels [38]. Likewise, eutrophication-mediated increase in organic matter production and remineralization stimulates microbial mercury methylation which results in bioaccumulation of neurotoxic monomethyl mercury in phytoplankton [39]. Metals like cadmium, nickel and copper are shown to make coastal areas more heterotrophic by increasing ecosystem respiration [11, 40, 41]. However, some trace metals such as iron, silicon and copper could stimulate primary production in eutrophic coastal and estuarine waters by increasing the cellular uptake rates of macronutrients [42, 43]. Sometimes, the mechanisms employed by algal species to deal with metal and nutrient contamination can have negative consequences for other species or for the entire ecosystem. For example, the diatom *Pseudo-nitzschia* produces toxic domoic acid in response to copper-stressed conditions in nutrient-rich coastal waters, which is linked to amnesic shellfish poisoning [44]. Regardless of the intensification or reduction in primary production or adaptation of a defence mechanism, the synergistic effects of metal and nutrients on phytoplankton are significant; and as autotrophs constitute the base of coastal food webs, changes in their growth, defense or community-structure dynamics can greatly impact organisms at higher trophic levels and the flow of energy and matter.

Eutrophication and metal pollution mediate other stressors too, such as hypoxia. In large parts of the coastal ocean, eutrophication and metal pollution lead to hypoxia, which in turn stimulate metal eco-toxicity. For example, hypoxia increases the bioavailability of manganese in sediment, potentially increasing its toxicity for pelagic and benthic organisms [45, 46].

## 4.3 | Mollusks

Mollusks are the second most studied taxonomic group after phytoplankton in the multistressor literature and the most critical interactive stressor for mollusks is ocean acidification combined with climate heating. The sensitivity to acidification is a consequence of the production of calcium carbonate shells[47]; the sensitivity to heating is metabolic stimulation with higher temperature, up to a critical threshold ($CT_{max}$) at which physiological process rates start to decline again [48, 49, 50]. [17] suggested that heating affects the larval development, whereas acidification would affect the reproductive capacity of adults.

A moderate increase in water temperature, can counteract the growth effects of reduced pH in *M. galloprovincialis* by allowing more active feeding time [51], and thus constituting an antagonistic interaction. Mostly, however, the combined effect of OA and heating in mollusks has been observed as negatively synergistic: Thomsen et al. [52], e.g., showed that heat shock proteins are downregulated under elevated $p\text{CO}_2$, amplifying heat stress experienced by *Mytilus edulis*. Many traits may be affected by the combination of heating and acidification: for both *M. edulis* and *Mytilus galloprovincialis* growth rate, shell size, and acid-base buffering capacity were found to decrease [53, 54].

The occurrence of further stressors along with heating and acidification exacerbates the negatively synergistic effects. Adding hypoxia impairs the fitness of marine mussels by reducing the activity of digestive enzymes [55]. Likewise, an increased frequency of extreme climatic events has been reported to impact bivalve mollusk species at 12 coastal regions around the Mediterranean [18].

Multistressor effects on mollusk species propagate to the entire coastal ecosystem via the ecosystem services provided, foremost the habitat creation and water quality improving filtration services, but also food provisioning. Presence of mollusks prevents the proliferation of harmful algal blooms [56]. Nutritional composition and thus commercial value of oysters and mussels decreases under combined acidification and heating stress [57].

## 4.4 | Research gaps

Effects of different stressor levels on populations in orthogonal experimental designs is difficult to examine, particularly when many stressors are involved [3]. It is costly in terms of time and resources and the inclusion of different stressor intensities may complicate the experimental design [58].

We identified only few studies (only 8%) which investigated the interactive effects of multiple stressors on ecosystem-level dynamics in coastal waters. This has been also noted by [2] in their review of non-additive effects of human stressors in marine systems, pointing out the existing bias towards studying single species in ex-situ setups. Experimental determination of complex interactions in coastal environments is challenging [5], and it is difficult to measure stressor effects at the community or ecosystem level in natural settings [59, 60, 61, 62].

As the interactive effects at every trophic level vary, depending on factors such as stressor magnitude and exposure duration, measurements of multiple endpoints have to be considered while designing the experiments on an ecosystem scale. [58]. A first step would start with individual stressor studies across a wide range of intensities to understand responses and then examining combinations of multiple stressors across a smaller range of stressor levels to explore interactions between them. This approach would help to identify critical stressors for multiple stressor experiments. Another difficulty in studying interactive effects across ecosystems is the variable response time of different taxa to a variety of stressors [3, 63]. Furthermore, the time at which stressor response is measured can also affect the classification of the interaction type [64].

Despite the complications in measuring ecosystem responses to multiple stressors, some efforts have been made in this direction. With the advent of mesocosm experiments, it has become possible to study ecosystem-level responses and effects of climatic and anthropogenic stressors in quasi-natural habitats [65]. For example, the mesocosm experiments of [66] facilitated assessment for a whole range of effects of multiple stressors (temperature, salinity, pH, light) on benthic ecosystems and communities. Given the large number of stressor combinations and species in coastal ecosystems, it will remain unfeasible to fully understand multi-factorial stressor effects by means of observational experiments. Therefore, other approaches such as modelling and expert opinion have been proposed [1, 3]. For modelling, a 3-tiered approach has been proposed which includes first mechanistic understanding of the stressor effects at the individual level [3], and then to scale these to population-level responses and, finally, to assess the risks for communities across ecosystems. This practice has been adopted by some modelling studies [67, 68, 69].

Although we found that the majority of multi-stressor studies focus on species-level effects, these are largely restricted to phytoplankton and mollusks. Other taxonomic groups, such as zooplankton, fish and benthic organisms, which constitute important trophic linkages in the coastal food webs, are underrepresented (Figs. 4 and 5). It is difficult to manage physiologically complex and larger organisms in manipulative experiments and to track variable responses to the same stressors by different species of the same group, e.g. in fish. Most of the multi-stressor studies performed on seagrass are in-situ experiments where co-variations in environmental boundary conditions make it hard to unravel non-additive interactions [70]. As a consequence, we could not identify a trend in reported interactive effects for seagrass or fish. Intra-specific responses in higher organisms (such as fish) vary depending on stressor magnitude and duration [71]. Stressor-driven changes at higher trophic levels trigger cascading effects that impact the food web dynamics in coastal ecosystems [72, 73, 74]. The same applies to the coupling of pelagic and benthic ecosystems. In our analysis, we found that only a few studies focussed on benthic polychaetes and those who did mostly reported synergistic effects (having negative consequences) in response to multiple stressors. Benthic poly-chaetes are powerful ecosystem engineers as they contribute to nutrient recycling, carbon storage and oxygenation of sediments. [75, 76].

Thus, the holistic understanding of how coastal ecosystems will evolve with ongoing changes requires knowledge on how different trophic levels will respond to multiple stressors. Therefore, it is important to diversify the target taxonomic group in multi-stressor studies or experiments. Our review also disclosed that relevant stressors such as hypoxia, turbidity, and invasive species, are understudied. Hypoxia has been identified as one of the most critical stressors for coastal ecosystems, negatively affecting pelagic and benthic organisms [77]. Other stressors such as warming, eutrophication and ocean acidification create feedback loops with hypoxia and together they intensify the stress on coastal food webs [78, 79]. Hypoxia, particularly, increases respiratory stress in benthic fauna (such as arthropods and mollusks) and in fish, which can trigger cascading effects [80, 81, 82]. Likewise, fishing, introduction of invasive species and turbidity have been reported to alter trophodynamics [73, 83, 84]. Thus, more experimental and modelling efforts should be put together to investigate the non-additive effects of these stressors at species and community levels in coastal ecosystems.

## 4.5 | Acclimation and adaptation strategies to stressors in coastal oceans

In coastal and open oceans, stressor effects are mitigated by acclimation strategies employed by organism [85, 86, 87]. Generally, acclimation is manifested in the form of trade-offs in trait responses, which becomes adaptation when this information is genetically passed on to the offspring. To cope with stressors, organisms modulate their physiology, morphology and behaviour. However, the mode of response varies between taxonomic groups.

In marine environments, phytoplankton possess high phenotypic plasticity [85]. For example, autotrophs may compensate for the losses associated with nutrient limitation, owing to climate warming, by the gains from concomitant ocean acidification [88]. High $CO_2$ and warm conditions allow some phytoplankton (such as *Ulva linza*) to reduce the metabolic and photosynthetic costs and reallocate the saved energy for nutrient acquisition which gives them a competitive edge over other algal species in coastal waters [89]. High-temperature adapted diatom species show higher tolerance to metal pollution[90, 91]. Likewise, thermally-acclimated species of some coastal diatoms showed reduced photo-inhibition during exposure to high light compared to the others as the negative growth rates due to UV inhibition were compensated by temperature-driven stimulation in non-photochemical quenching [92, 93]. Similarly, high-temperature adapted coccolithophores showed higher fitness under elevated $p$CO$_2$ levels compared to non-adapted ones [94].

Besides phytoplankton, other taxonomic groups compensate for negative stressor effects as well through, trait-mediated acclimation and adaptation. Zooplankton responds to stressor-driven environmental changes (such as climate warming, hypoxia and pollutants) by phenotypic plasticity and genetic adaptation [86]. Zooplankton can adapt to temperature changes but there is a latitudinal gradient in the thermal-stress tolerance capacity [95]. As the rate of warming in high latitudes of the Northern Hemisphere is twice the global mean, zooplankton species in temperate regions show higher adaptive capacity to temperature increase compared to their tropical counterparts [96]. However, this adaptation comes with costs. A decrease in zooplankton body size has been observed with the warming in shelf seas and coastal regions [97]. Zooplankton also adapts to toxins in coastal oceans [86, 98, 99]. Grazers with exposure to pollutants are adapted to deal with toxic food (toxin-producing algae) compared to non-adapted populations [100]. Likewise, zooplankton in coastal environments can also adapt to hypoxia. Small crustaceans such as copepods and euphausiids are shown to tolerate much lower dissolved oxygen levels than larger zooplankton [101]. They do so through physiological adaptions (e.g. by having large gill surfaces, short diffusion distances and by activating respiratory proteins with very high oxygen affinity) or by behavioural adaption such as vertical migration [102, 103].

Bivalve species have been reported to exhibit long-term adaptation to ocean acidification, salinity, hypoxia and temperature, especially if these have wide thermal and thermohaline tolerance limits [104, 105, 106]. This adaptation can be a physiological one, such as such as upregulation of proteins involved in calcification, metabolic processes, oxidative stress and high thermal tolerance [107, 108, 109]. Physiological, but also behavioural acclimation to multiple stressors has been reported in fish as well [110, 111, 112]. Prior exposure to heat stress leads to improvement in hypoxia tolerance in coastal demersal fish *Oligocottus maculosus*, *Fundulus heteroclitus* and *Zoarces viviparus*[113, 114, 115]. Likewise, acclimation to oxygen stress results in an increase in thermal tolerance in channel catfish, *Ictalurus punctatus* [116]. Adaptation to multiple stressors (such as temperature, salinity, oxygen, and anthropogenic toxicants) has been reported in common killifish, *Fundulus heteroclitus* [117].

## 5 | CONCLUSION

In our review analysis, we identified climate heating, eutrophication, ocean acidification and metal pollution, or combinations thereof, as the most critical stressors in triggering non-additive responses in coastal ecosystems. Phytoplankton and mollusks are the most studied taxonomic groups in multi-stressor experiments and observations. They are, however, sensitive to different stressor combinations. Simultaneous exposure to metal pollution and high nutrient concentrations invokes synergistic responses in phytoplankton, often with negative effects on their physiology. For mollusks, climate heating and ocean acidification appear to be the most critical stressor combination, with adverse effects on physiological and morphological traits. Organisms at different trophic levels, or belonging to different ecosystem components (such as micro- and macrobenthos), are sensitive to different stressor combinations. Consequently, interactive effects could induce radical changes in trophodynamics and food web structure as climate and human-induced changes in coastal ecosystems continue to intensify. For a holistic understanding of cumulative effects of multiple stressors in coastal ecosystems, however, we suggest that more research focus should be placed on studying community and ecosystem-level responses. This can be achieved by combining in-situ observations (e.g. mesocosm experiments) with modelling approaches. In addition, other relevant taxonomic groups such as zooplankton, fish and benthic polychaetes demand more research.

| Study | Stressors | Organisms | Synergistic (Yes/No) | Antagonistic (Yes/No) | Additive (Yes/No) |
|---|---|---|---|---|---|
| e.g. ABC et al., 2016 | pH,temp | *M.edulis* | Yes | No | No |
| e.g. XYZ et al., 2010 | metal, nut | *T. pseudonana* | Yes | Yes | No |

**TABLE 1**    The format of the Summary Table which was provided to all co-authors to include meta data information. The information shown here is just an example. The table contained other relevant information as well, which is not listed above, such as study-location, study-type, time-scale and traits investigated.

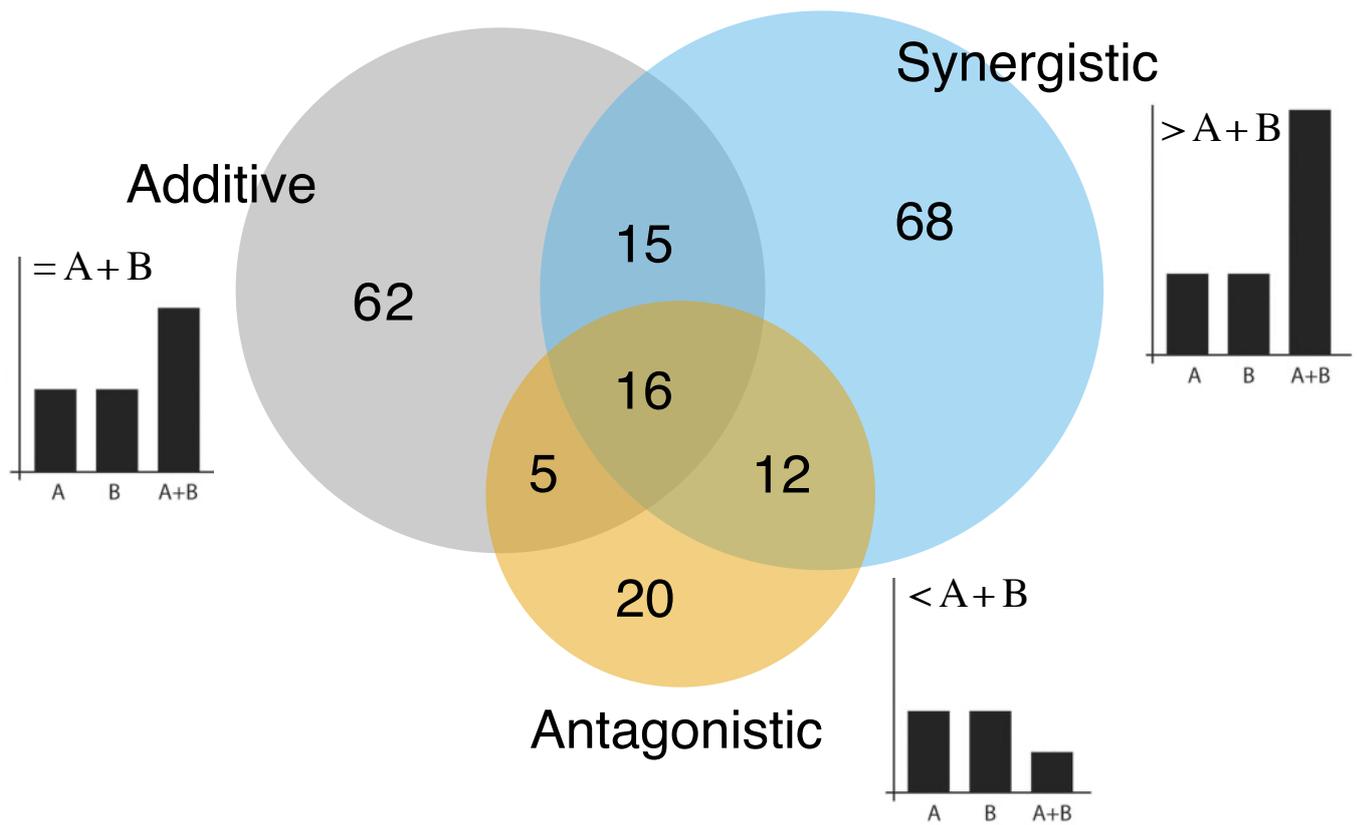

**FIGURE 1** Venn diagram showing the number of studies which reported Synergistic, Antagonistic and Additive effects in coastal ecosystems (across taxonomic groups). The overlaps show the number of studies reporting two or more than two interactive effects.

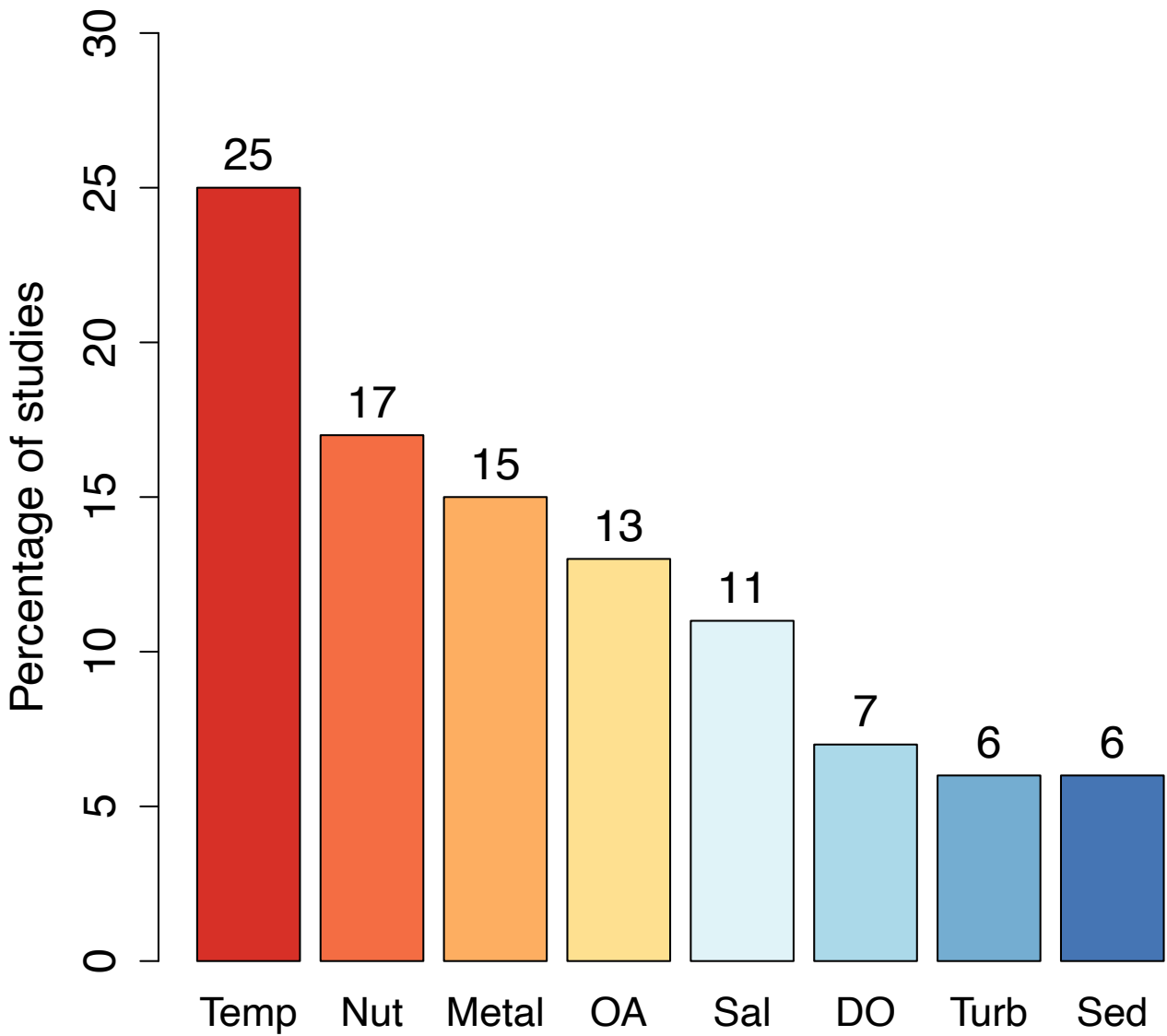

**FIGURE 2** Most studied stressors in coastal ecosystems (in terms of percentage), identified from our review. Temp = Temperature/warming, Nut = Nutrient pollution/Eutrophication, Metal = Metal pollution, OA = Ocean acidification, Sal = Salinity, DO = Dissolved oxygen, Turb = Turbidity, Sed = Sedimentation

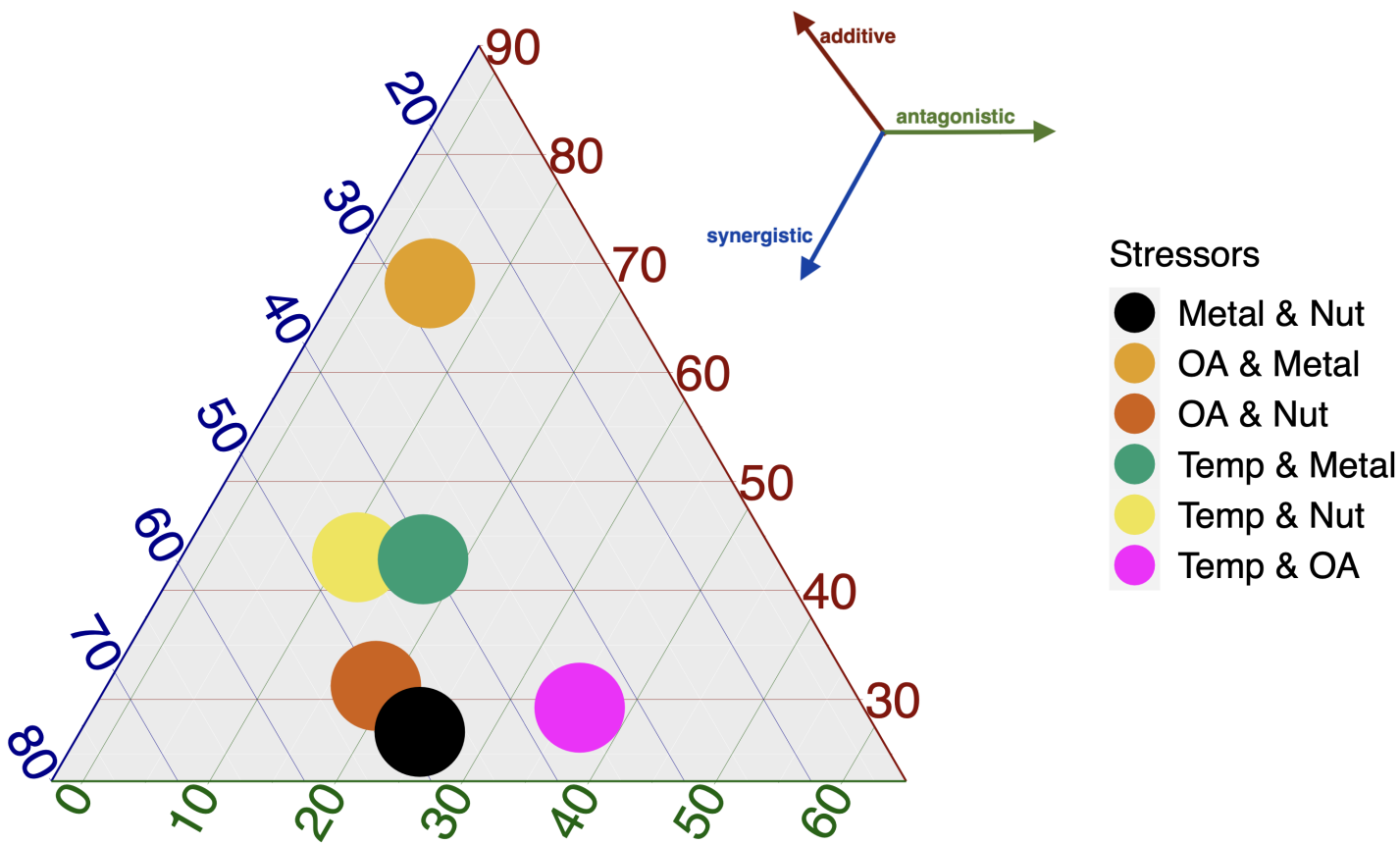

**FIGURE 3** The above ternary plot shows the percentage score for the reported interactive effect ("Syn" = Synergistic, "Ant" = Antagonistic, "Add" = Additive) corresponding to the different stressor combinations (Metal & Nut, OA & Metal, OA & Nut, Temp & Metal, Temp & Nut, Temp & OA) that are indicated by the colours of solid circles. The score (0 to 100 %) for the synergistic effect increases towards the left vertex of the triangle as indicated by the blue corner lines. Likewise, the scores for the antagonistic (the green lines) and the additive (the red corner lines) effects increase towards the right and top vertexes, respectively. For an explanation of the stressor abbreviations see Figure 2.

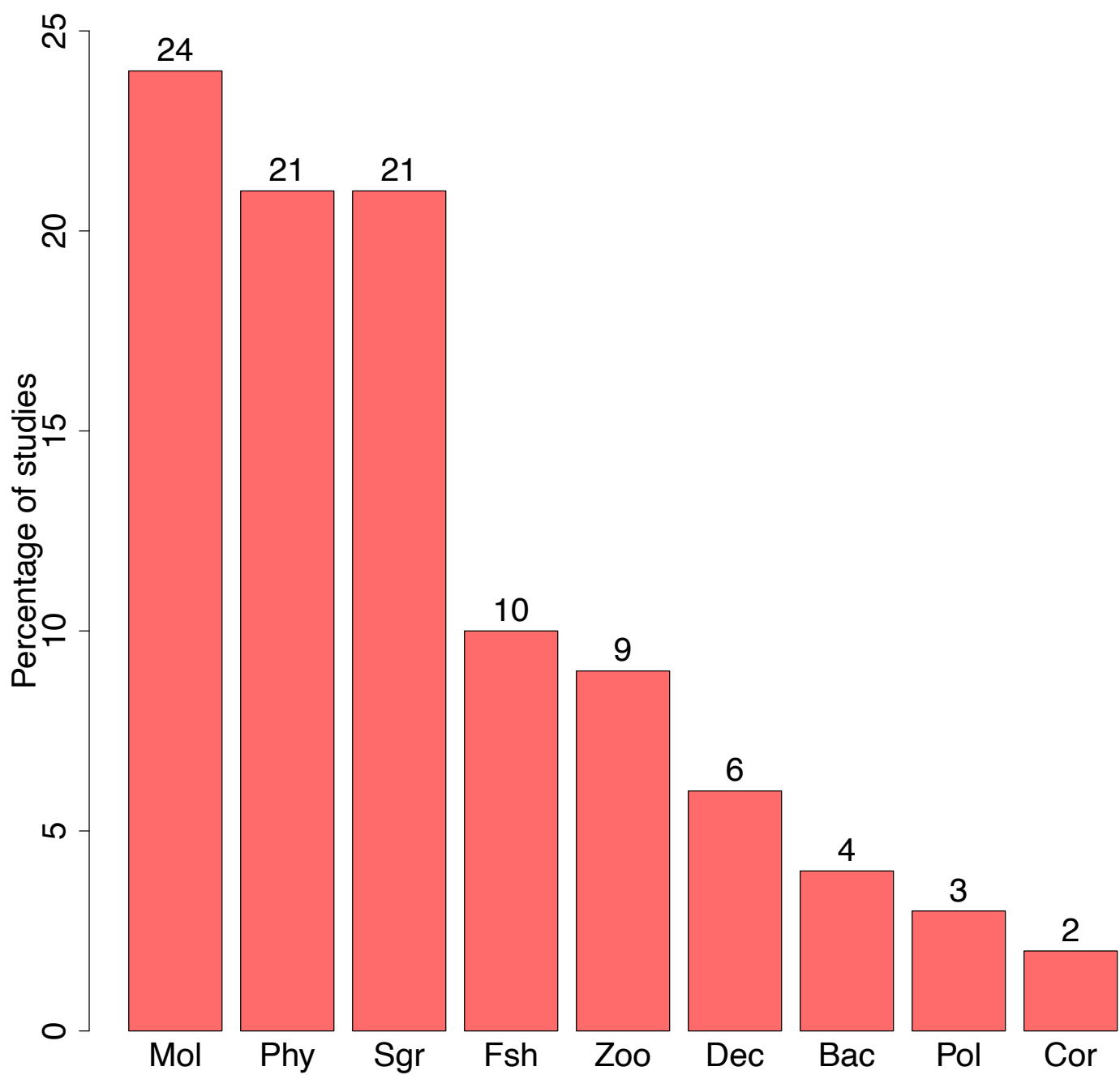

**FIGURE 4** The most studied taxonomic groups (in terms of percentage) with respect to multi-stressor effects in coastal ecosystems. Mol = Mollusks, Phy = Phytoplankton, Sgr = Seagrass, Fsh = Fish, Zoo = Zooplankton, Dec = Decapods, Bac = Bacteria, Pol = Polychaete, Cor = Corals.

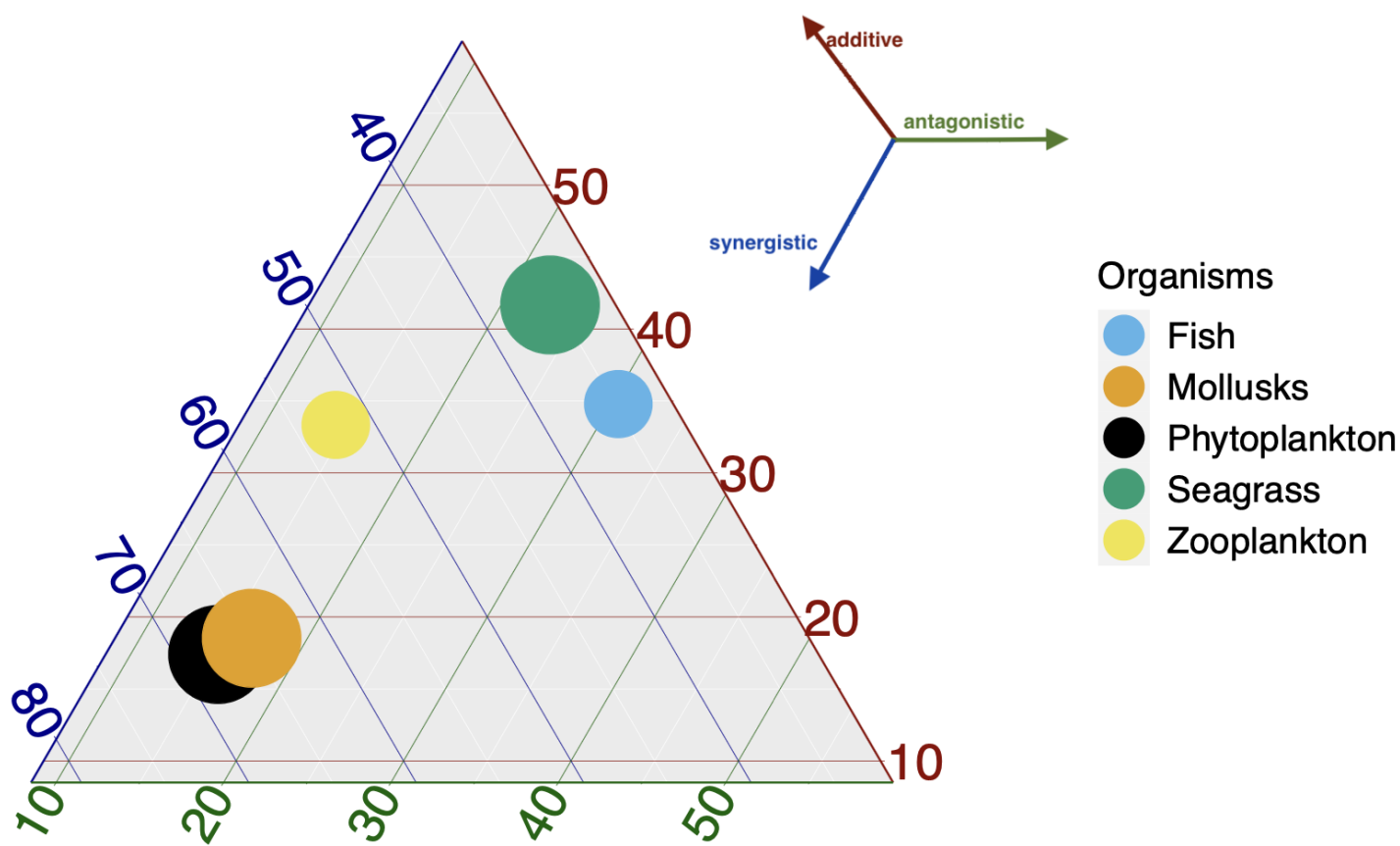

**FIGURE 5** The above ternary plot shows the percentage score for the reported interactive effect ("Syn" = Synergistic, "Ant" = Antagonistic, "Add" = Additive) in different taxonomic groups highlighted by different colours. The score ( 0 to 100 %) for the synergistic effect increases towards the left vertex of the triangle as indicated by the blue corner lines. Likewise, the scores for the antagonistic (the green lines) and the additive (the red corner lines) effects increase towards the right and top vertexes, respectively. The solid circles show the score for a given stressor combination. The size of the circles indicates the sample size of the studies.

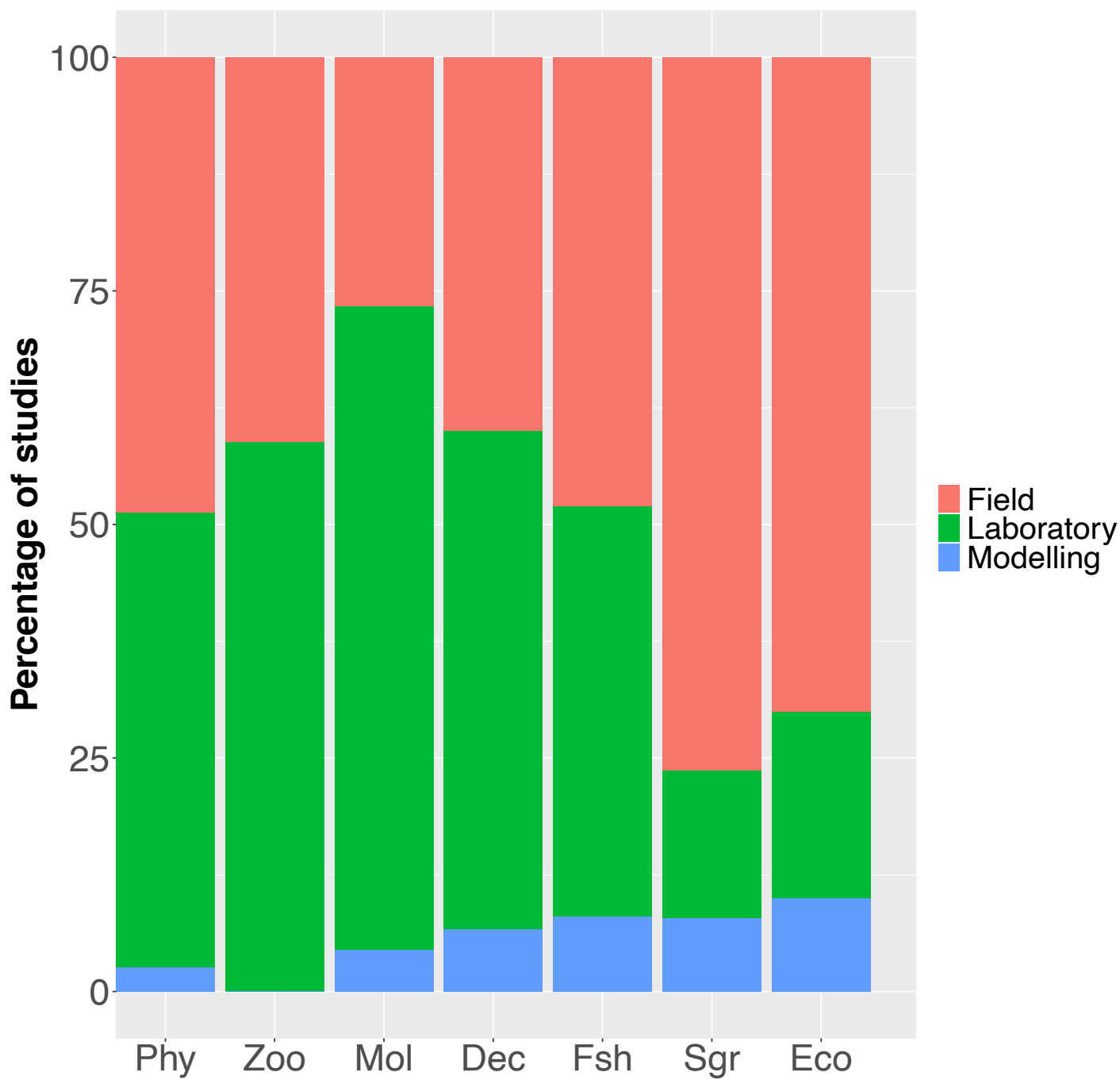

**FIGURE 6** Study types (field, laboratory, and modelling) expressed in terms of percentage corresponding to different taxonomic groups (of increasing complexity) and to ecosystems. Phy = Phytoplankton, Zoo = Zooplankton, Mol = Mollusks, Dec = Decapods, Fsh = Fish, Sgr = Seagrass, Eco = Ecosystem.

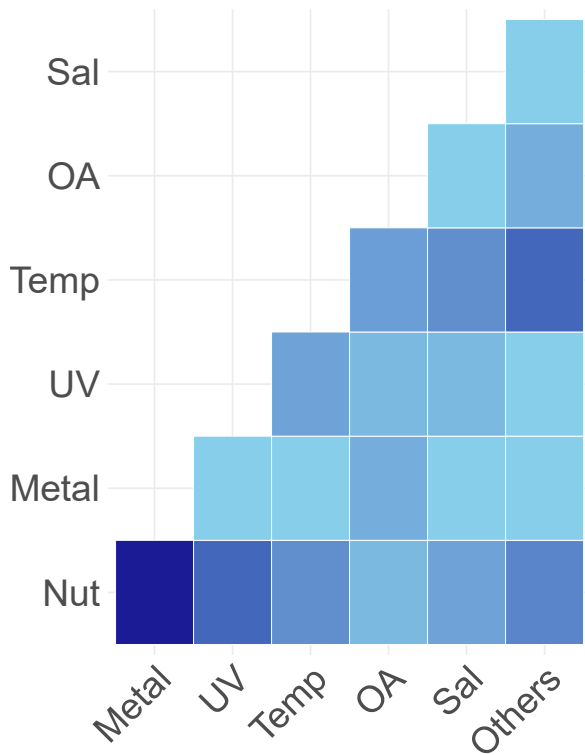
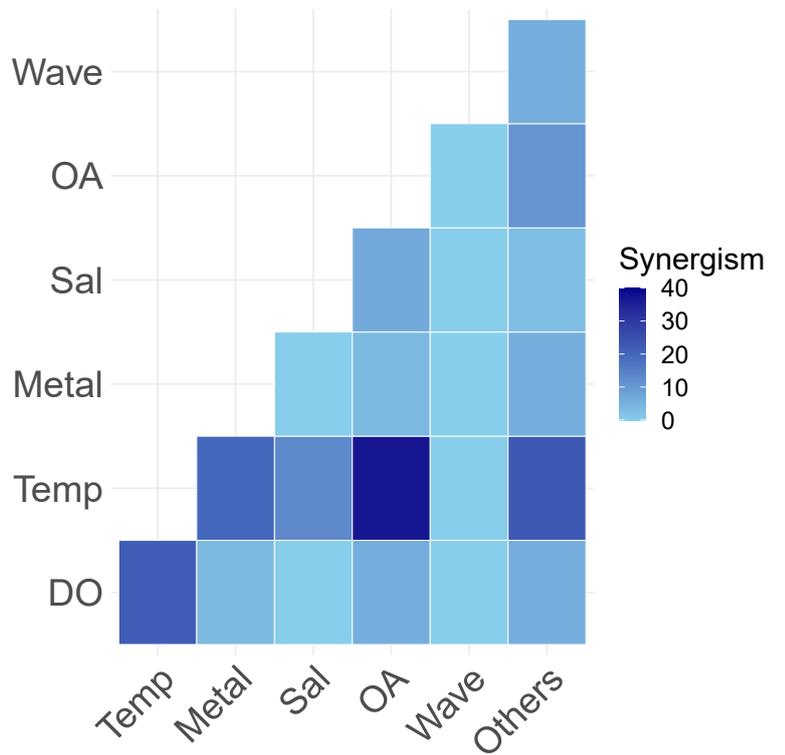

**FIGURE 7** Critical stressor combination for phytoplankton (A) and Mollusks (B) based on stressor-pair scores. Dark blue tone indicates a high score for synergism and light blue a low score. For an explanation of the stressor abbreviations see Figure 2.

## Acknowledgements

We are grateful to Cedric Meunier and Ingrid Kroencke for their valuable input in the development of this paper.

## Conflict of interest

No conflict of interest.

## Supporting information

Additional figures.

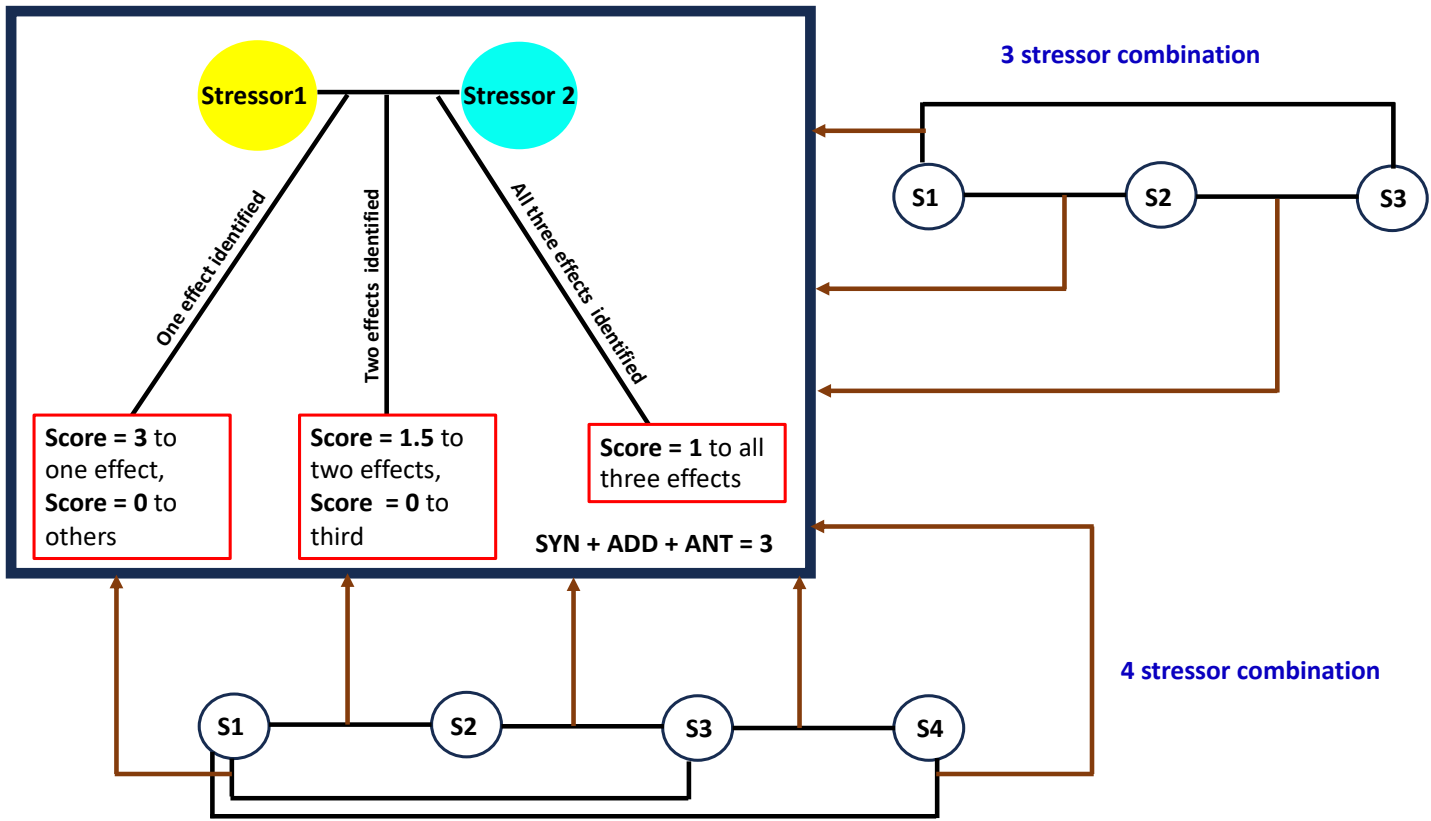

**FIGURE 8** Schematic of score distributions for the synergistic (SYN), antagonistic (ANT), and additive (ADD) effects corresponding to stressor pair (black box). The sum of SYN, ADD, and ADD effects is always 3 for a given stressor pair. The brown arrows show the stressor pairs resulting from 3 and 4 stressor combinations.

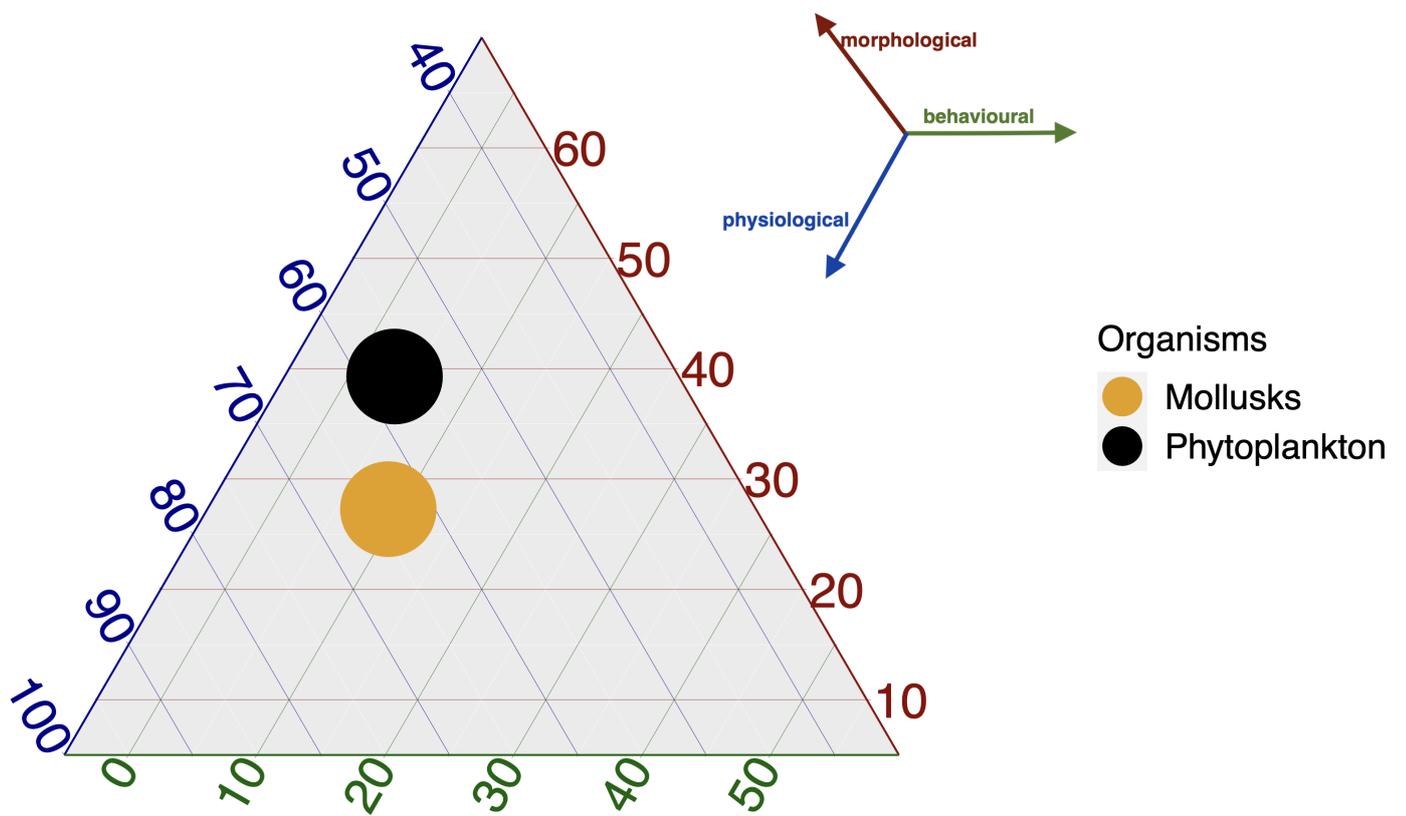

**FIGURE 9** Ternary plots showing the percentage score for the type of studied traits (Physiological, Behavioural, and Morphological) in Phytoplankton and in Mollusks.

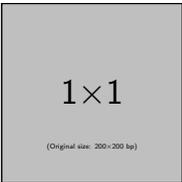

**A. One** Please check with the journal's author guidelines whether author biographies are required. They are usually only included for review-type articles, and typically require photos and brief biographies (up to 75 words) for each author.

## GRAPHICAL ABSTRACT

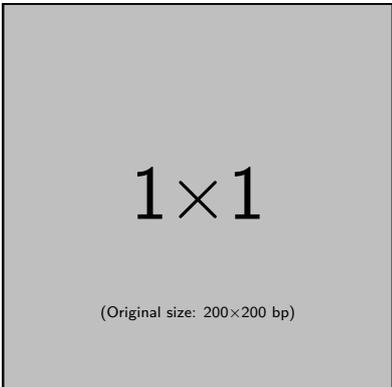

Please check the journal's author guildines for whether a graphical abstract, key points, new findings, or other items are required for display in the Table of Contents.